\documentclass[runningheads]{llncs}
\usepackage{graphicx} 
\usepackage{algorithm}
\usepackage{algorithmic}
\usepackage{multirow}
\usepackage{enumitem}
\usepackage{subfigure}
\usepackage[T1]{fontenc}
\begin{document}
\title{ Integrating Explainable AI for Effective Malware Detection in Encrypted Network Traffic}

 \titlerunning{Explainable AI for Encrypted Malware Traffic}

\author{Sileshi Nibret Zeleke\inst{1} \orcidID{0009-0006-8172-9646}
Amsalu Fentie Jember\inst{1}
 \orcidID{0009-0004-7356-682X} 
,\and
Mario Bochicchio\inst{1,2}
 \orcidID{0000-0002-9122-6317}
}

\authorrunning{Sileshi N. Zeleke et al.}
\institute{Department of Computer Science, University of Bari, Bari, Italy 
\and Digital Health National Lab, CINI - Consorzio Interuniversitario Nazionale per l’Informatica, Roma, Italy
\email{\{sileshi.zeleke,amsalu.jember,mario.bochicchio\}@uniba.it}} 

\maketitle     

\begin{abstract}  
Encrypted network communication ensures confidentiality, integrity, and privacy between endpoints. However, attackers are increasingly exploiting encryption to conceal malicious behavior. Detecting unknown encrypted malicious traffic without decrypting the payloads remains a significant challenge. In this study, we investigate the integration of explainable artificial intelligence (XAI) techniques to detect malicious network traffic. We employ ensemble learning models to identify malicious activity using multi-view features extracted from various aspects of encrypted communication. To effectively represent malicious communication, we compiled a robust dataset with 1,127 unique connections, more than any other available open-source dataset, and spanning 54 malware families. Our models were benchmarked against the CTU-13 dataset, achieving performance of over 99\% accuracy, precision, and F1-score. Additionally, the eXtreme Gradient Boosting (XGB) model demonstrated 99.32\% accuracy, 99.53\% precision, and 99.43\% F1-score on our custom dataset. By leveraging Shapley Additive Explanations (SHAP), we identified that the maximum packet size, mean inter-arrival time of packets, and transport layer security version used are the most critical features for the global model explanation. Furthermore, key features were identified as important for local explanations across both datasets for individual traffic samples. These insights provide a deeper understanding of the model's decision-making process, enhancing the transparency and reliability of detecting malicious encrypted traffic.

\end{abstract}

\section{Introduction}

Malware represents a persistent cyber threat, and with the widespread adoption of encryption in network communications, it is increasingly being transmitted over encrypted channels \cite{32}. As encryption becomes more prevalent for securing networks, including its use by malware to evade detection and analysis, the ability to inspect encrypted traffic for signs of malicious behavior has become crucial for effective cybersecurity. This gap is addressed by encrypted network analysis, which provides methods and resources for deciphering encrypted data and identifying indicators of compromise or hostile activity over encrypted communication channels. Traditional malware detection techniques may struggle to identify encrypted malware, and deep packet inspection can compromise the privacy of the payload \cite{35}. To overcome this challenge, some studies have focused on analyzing encrypted traffic through decryption, which can also compromise privacy. However, many of these studies primarily consider the efficiency or accuracy of intelligence-based detection systems. Most research prioritizes the efficiency and accuracy of these systems while neglecting the importance of explainability in cybersecurity tasks. A better understanding of the model's decisions in classifying specific traffic flows as malicious or normal is essential for effective analytics. 

Explainable Artificial Intelligence (XAI) has emerged as a means to interpret the outputs of machine learning algorithms. XAI techniques can be categorized into local and global interpretations. Local methods provide insights into specific instances, while global methods offer a comprehensive interpretation of the model. Global explanations focus on interpreting the model's behavior across the entire test sample, enhancing the interpretability of the model. XAI techniques can be either model-agnostic or model-dependent, depending on how much they rely on specific AI models. While model-agnostic approaches can theoretically be applied to any AI model, model-dependent methods are specifically designed for a given model \cite{34}. 

In this study, We analyzed raw encrypted traffic from 6 different sources to
extract and analyze different feature sets that can discriminate malicious flows
from normal flows. The features include a handshake, certificate, inter-arrival time and packet length, statistical features, meta-connection features, and cipher suite used. Despite progress, challenges remain in multi-view feature analysis, deep feature analysis, and efficient detection and classification models. We extract diverse features from encrypted traffic and regard encrypted traffic as sample nodes, then use AI for malicious traffic detection. This research addresses gaps in explainability of model decisions and feature engineering of flows of encrypted network traffic. A diverse feature set of encrypted network communication, including server-side features, have important roles in detection but have not been analyzed in previous studies.

The main contributions of this study are:
\begin{itemize}[label=$-$]
 \item Explainable approach to detect encrypted malware in network traffic.
\item Identify important features from server-side, time-related, payload-related, or other sets for detection models.
\item A dataset containing 1,127 malware traffic from 54 different malware families which represent significant number of malware variants is presented for the research community.  \end{itemize}
The rest of the paper is organized as follows. Section 2, is a literature review of existing encrypted malware detection studies and usage of explainable models for malware detection. Section 3 describes the proposed malware detection methodology. The experimental setup and test results are presented in Section 4. However, Section 5 discusses and concludes the paper with remarks and future work ideas.

\section {Related Work}
  
Recent advances in encrypted traffic analysis for malware detection and XAI have attracted significant research attention. Notably, Cisco has pioneered the implementation of an encrypted traffic analyzer in their security devices. The proposed supervised models are trained with multi-view features such as transport layer security (TLS) handshake metadata. Moreover, DNS contextual flows and HTTP metadata from the same source IP within a 5-minute interval, have shown the potential of AI models to identify malware traffic \cite{3}. By observing the disparities between malicious and normal network flow’s contextual information were can capture strong discriminatory feature set \cite{4}. 

Analytical techniques to infer HTTP semantics from passive observations of HTTPS can reveal the significance of important fields \cite{5}. The study conducted in \cite{6} performed a feature analysis of encrypted malicious traffic within HTTPS network traffic, utilizing datasets captured from two research projects at the Czech Technical University. This analysis involved 72 network traffic captures, comprising 59 malware and 13 normal pcap files. Another study \cite{7} identified encrypted malware traffic using unsupervised learning methods. The distance metric was employed to construct a new malware class termed FClass. 83 numerical features were extracted from four categories: TCP/IP header, time-based, length-related, and packet variation features. The 10 most relevant features were then utilized for classification. In the study presented in \cite{8}, NetConverse examined the flow of traffic associated with various ransomware families within the Windows ransomware network. Features were extracted from unencrypted web traffic to train classifiers, achieving an accuracy of 97.1\% with a decision tree classifier. A primary limitation of this paper is that it extracts features solely from unencrypted traffic and focuses on only a subset of ransomware families. Biflow, as described in \cite{9}, is a flow-based system that aggregates data packets from two families.

In addition, the aforementioned experiments were conducted on a limited number of ransomware families. To enhance the accuracy of malware identification, a technique proposed by \cite{13} involves decrypting suspected encrypted flows and performing conventional deep packet inspection using intrusion signatures. Another study by \cite{14} utilized statistical and sequence features from both flow-level and host-level perspectives to characterize encrypted traffic, acknowledging the challenges associated with decryption. They introduced a detection framework based on ensemble learning that incorporates real malware, including normal traffic generated by legitimate hosts and malicious traffic produced by malware-infected hosts.

Most studies have focused on the availability of expert-labeled data for detection algorithms. However, the study by \cite{15} addressed the challenge posed by the scarcity of high-dimensional labeled data by proposing an unsupervised anomaly detection method. This method utilizes a three-layer autoencoder for feature compression, enhancing model efficiency, and employs the classical K-means algorithm for unsupervised classification. In addition to the proposed solution, the study exclusively analyzed normal encrypted traffic. To identify encrypted malicious traffic, Wang et al. \cite{32} extracted characteristics from HTTPS traffic and integrated them into shallow machine-learning models. To tackle the issue of low classification accuracy in current encrypted traffic classification methods, particularly for traffic with similar fingerprints, Shen et al. \cite{11} developed an attribute-aware encrypted traffic classification method based on second-order Markov chains.

In general, multi-view feature extraction, machine learning algorithms, and ensemble learning frameworks play a critical role in malware detection within encrypted networks. While most studies emphasize enhancing the accuracy and efficiency of detecting malware in encrypted traffic, the explainability of AI models remains a crucial aspect that requires further investigation.

\section{Methodology} 
\begin{figure}
\centering
\includegraphics[width=12cm, height=8cm]{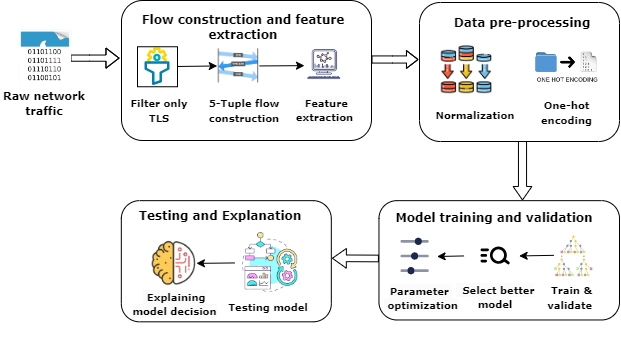}
\caption{Overview of the proposed explainable malware detection pipeline } 
\label{fig1}
\label{general_arch}
\end{figure}

The overall design of the proposed framework, as illustrated in Fig. \ref{general_arch}, consists of four core components: flow construction and feature extraction, data preprocessing, model training and validation, and the explanation module. Below, we provide a detailed description of each component. These components highlight their synergistic interactions within the framework. To the best of our knowledge, an explainable encrypted malware detection system at the network level has not been studied. This makes our proposed method a novel approach, as it integrates flow construction, feature extraction, data preprocessing, and XAI.

\subsection{Data Preparation}
We collect publicly available and proprietary malicious network traffic to create a robust dataset that encompasses various types and families of malware. To achieve this, we gathered raw network traffic from multiple sources as follows: We obtained 614 ransomware traffic samples from 17 different ransomware types provided by the Information Security and Object Lab. Additionally, we acquired more Trojan horse samples from \cite{36}, which includes 175 raw network traffic samples composed of 12 families, the sample includes the notorious Zeus and Emotet. Another source of malware traffic for our analysis is malware traffic analysis \cite{17} and the Czech Republic (CTU-13) \cite{18}, from which we obtained 305 and 36 traffic samples, respectively. The malicious samples collected from the aforementioned sources comprise 54 different families, each exhibiting distinct communication properties, as shown in Table \ref{malware_family_total}. The sample count varies among families due to the availability of raw traffic. For example, Teslacrypt is the most represented malware family in our dataset. For the machine learning model used for malware detection, the dataset is not problematic even if the families are imbalanced, as all malware families represent a single class (i.e., the malware class).  

Acquiring normal traffic captured in a local area network (LAN) setup at the enterprise network edge or host is relatively straightforward compared to obtaining malware traffic. In our study, we utilized two sources for normal traffic data: self-collected traffic from the Addis Ababa Science and Technology campus network and data from the CTU-13 dataset \cite{18}. For the self-collected data, we configured a local host to capture traffic flow while visiting web servers of legitimate Fortune companies. The traffic collection procedure is briefly presented on \cite{31}. The comprehensive data processing procedure is presented in detail in Algorithm 1.
\subsection{Flow Construction and Feature Extraction} 
To better characterize the communication between the client and server, it is essential to establish a bi-directional flow based on the source and destination ports, source and destination IP addresses, and the connection protocol. We utilized Joy framework \cite{19}, an open-source tool for flow generation, to create flows from raw network traffic. After constructing a 5-tuple bi-directional flow, we proceeded to extract features from various perspectives. In our multi-view feature extraction strategy, we extract features from connection metadata, certificate information, time and length-related data, handshake details, and statistical features. Connection flow metadata describes the number of bytes or packets sent to or received from the client within a specific time window; in our case, this window is 30 minutes. The packet length and inter-arrival time between two consecutive sessions vary for normal and malware communications \cite{20}. 

To represent these features, we employed one-hot encoding, such that if a certain cipher suite is present in the offer or accepted list, we encode it as 1; otherwise, we encode it as 0. The same technique is applied to TLS extensions and version numbers. To extract time and packet-related features, we utilized a Markov chain sequencing by discretizing the data into equal-sized bins. A Markov chain consists of a set of states and transitions sequentially from one state to another \cite{37}.
 \begin{table}[H] 
\caption{Malware family, type, and number of samples included as a raw pcap}
\label{malware_family_total}
\centering
\begin{tabular}{l|l|c|l|l|c}
\hline
Family &  Type  &  \# Samples  &  Family & Type      &  \# Samples\\ \hline
Cerber                & Ransomware               & 124                     & Win32.Blocker         & Ransomware    & 18                      \\  
Mole                  & Ransomware               & 4                       & Zeta                  & Ransomware               & 3                       \\ 
CryptoShield          & Ransomware               & 2                       & Zeus                  & Trojan                   & 101                     \\  

Jaff                  & Ransomware               & 3                       & Nemucod               & Ransomware               & 2                       \\  
Unlock26              & Ransomware               & 3                       & Stealer               & Ransomware               & 3                       \\  
Locky                 & Ransomware               & 100                     & BankerX-gen           & Ransomware               & 3                       \\  
WannaCry              & Ransomware               & 10                      & Rig                   & Downloader               & 17                      \\ 
Xorist                & Ransomware               & 2                       & Trickbot              & Trojan           & 88                      \\  
Zeus-panda            &  Trojan           & 23                      & Icedid                & Trojan           & 13                      \\  
Gootkit               &  Trojan           & 13                      & Ursnif                & Virus                    & 5                       \\  
Neutrino              & Rootkits                 & 2                       & Boleto                & Downloader               & 12                      \\  
ZLoader               & Ransomware               & 2                       & Vawtrak               & Trojan           & 7                       \\  
BazarLoad             & Backdoor                 & 1                       & Kovter                & Trojan                   & 3                       \\  
Hancitor              & Rootkits                 & 54                      & Dreambot              & Trojan                   & 9                       \\  
Nymaim                & Downloader               & 1                       & Ramnit                & Worm                     & 1                       \\  
Petya                 & Ransomware               & 2                       & Upatre                & Ransomware               & 10                      \\  
Crysis                & Ransomware               & 8                       & Crypt R               & Ransomware               & 14                      \\ 
Sage                  & Ransomware               & 5                       & Bunitu T              & Ransomware               & 10                      \\ 
CTBLocker             & Ransomware               & 2                       & Crthrazy              & Ransomware               & 6                       \\  
Spora                 & Ransomware               & 30                      & Troldesh              & Ransomware               & 9                       \\ 
GlobeImposter         & Ransomware               & 7                       & Emotet                & Ransomware               & 9                       \\ 
TeslaCrypt            & Ransomware               & 331                     & Tofsee                & Ransomware               & 11                      \\  
Dridex                & Trojan           & 23                      & Gandcrab              & Ransomware               & 3                       \\  
Azorult               & Spyware                  & 2                       & Loveyou               & Virus                    & 1                       \\  
Qakbot                &  Trojan           & 5                       & Spelevo               & Exploit Kit              & 3                       \\  
Chthonic              &  Trojan           & 1                       & Fallout               & Rootkits              & 1                       \\  
Angler                & Exploit Kit              & 2                       & Miuref                & Trojan       & 3                       \\  
\hline

\end{tabular}\vspace{-3mm}
\end{table}

 Moving through all states generates a transition matrix. We used three different states for our analysis. We experimented with two datasets for this study. The first dataset was self-compiled, as described in Sec. 3.1 and the other is the CTU-13. The CTU-13 botnet dataset contains 38,898 botnet samples and 53,314 normal samples. 
 
 \subsection{Detection Model}
To detect and classify malware in encrypted networks, various conventional and deep learning algorithms can be employed, as discussed in the existing literature. 

\begin{algorithm}[H]
\centering
\caption{Data Processing Algorithm }
\label{alg:data_processing}
\begin{algorithmic}[1]
\renewcommand{\algorithmicrequire}{\textbf{Input:}}
\renewcommand{\algorithmicensure}{\textbf{Output:}}
\REQUIRE Path of pcap files ($P$)
\ENSURE Normalized CSV file ($N$)
\STATE Initialize variables
\FOR{each $f \in P$}
    \STATE Construct 5-tuple flow ($F$)
    \IF{duration($F$) $<$ 30 min}
        \STATE Split $F$ into 30-min windows
        \STATE Filter non-encrypted flows
    \ELSE
        \STATE Discard $F$
    \ENDIF
    \IF{three-way handshake completed}
        \STATE Pre-process $F$
    \ELSE
        \STATE Discard $F$
    \ENDIF
    \IF{non-encrypted connection}
        \STATE Discard $F$
    \ELSE
        \FOR{each flow $f_i \in F$}
            \STATE Extract features (handshake, metadata, stats)
            \FOR{each packet $p_i$ and inter-arrival time $\tau_i$}
                \STATE Create 3 states (150 bytes/ms)
                \STATE Initialize $3 \times 3$ transition matrix ($M$)
                \STATE Determine state of $p_i$ in $M$
                \STATE Compute transition probabilities
            \ENDFOR
        \ENDFOR
        \STATE Update dataset features
    \ENDIF
    \STATE Append processed data to dataset
\ENDFOR
\RETURN $N$
\end{algorithmic}

\end{algorithm}
\vspace{-10pt}
However, studies \cite{21}, \cite{20}, \cite{6}, \cite{21}, \cite{22}, and \cite{23} indicate that tree-based ensemble models demonstrate superior performance. Specifically, study \cite{32} conducted a comparison of deep learning algorithms and random forest (RF) for analysis of encrypted malware traffic, concluding that RF outperformed all other models across every baseline presented. In light of this, we train and test ensemble algorithms, including RF, XGBoost, and extremely randomized trees.
\subsubsection{Random Forest:} Widely used ensemble learning method that operates on the principle of constructing a robust decision tree by aggregating the predictions of multiple trees, a technique also known as bagging.
\begin{equation}
y=\frac{1}{m} \sum_{i=1}^{m} \sum_{j=1}^{n} W_i(x_j,z)y_j
\end{equation}
Where \(y\) represents output, \(m\), \(n\),\(W\), and \(z\)  are a number of trees, data sample, weight value and new data point to be predicted respectively. Also \(i\) represents the respective tree and \(x\_j\) denotes the neighbor of \(z\) that share the same leaf in tree \(i\). To better exploit the performance of RF appropriate tuning of the number of estimators, maximum depth of the tree, minimum sample split, and minimum samples of leaf is important \cite{24}.

\subsubsection{Extreme Gradient Boosting:} Unlike RF, XGB employs a distinct technique known as boosting to create a robust classifier. This method incrementally enhances a weak classifier by adding one classifier at a time to improve the existing ensemble. According to \cite{25}, key parameters of XGB include the number of estimators, the maximum depth of the tree, learning rates, and the gamma value, which indicates the minimum loss reduction.
\subsubsection{Extremely Randomized Trees:} Also known as Extra Trees, has a structure similar to that of RF \cite{29}. However, the trees are generated with greater randomness to enhance diversity by selecting a random subset of features at each node. Additionally, this method employs random partitioning rather than seeking the optimal partition. This inherent randomness improves the model's generalization performance and helps prevent overfitting.

\subsection{Evaluation Metrics}
To evaluate performance, we considered precision, recall, F1-score, accuracy, and an evaluation metric known as the Matthews Correlation Coefficient (MCC). The MCC is particularly valuable because it incorporates all values from the confusion matrix to assess the effectiveness of a binary classification model, providing more comprehensive information than the F1-score and accuracy alone. This is primarily because the MCC accounts for the balance between true positives (TP), true negatives (TN), false positives (FP), and false negatives (FN).

\subsection{Explainability}  
To effectively explain tree-based models both globally and locally, SHAP has demonstrated strong performance in previous studies related to network traffic. SHAP is grounded in cooperative game theory, specifically the Shapley values \cite{27}. These values represent the average contribution of each player across all possible coalitions. In the context of machine learning, players are replaced by attributes from the sample, and their contributions are aggregated to determine the output of the algorithm. This technique falls under the category of additive feature attribution methods. SHAP is characterized by its properties of efficiency, symmetry, dummy, and additive. Regarding efficiency, the total sum of Shapley values, or the marginal contribution of each feature, is equal to the value of the total coalition. Symmetry ensures that each feature has an equal opportunity to influence the outcome, regardless of the order in which they are considered. If a particular feature does not affect the predicted value, regardless of the coalition group, its value is assigned as zero, which is referred to as the dummy property \cite{27}. Furthermore, we adopt TreeExplainer, a technique specifically designed for tree-based models \cite{28}.

\subsection{Experimental Settings}
The experiment was conducted on Google Colab Pro with V100 GPUs for both datasets. Moreover, To simulate the real-world scenario of malicious traffic occurrences in any given communication environment, we oversample our normal samples using an adaptive synthetic sampling technique. The normal-to-malware proportion of the imbalanced dataset is 98.97\% to 1.03\%. The other dataset is the CTU-13 botnet dataset \cite{18}. We used validation curves to identify optimal hyperparameters for the RF model. We train the model based on 10-fold cross-validation, the best parameters obtained from the validation curve are n\_estimators=23, max\_depth=42, min\_samples\_split=6, and min\_samples\_leaf=2. However, for XGB the best values of parameters were obtained using nature-inspired genetic algorithm, the best parameters are n\_estimators=23, max\_depth=43, learning\_rate=0.47, min\_child\_weight=0.4, gamma=3.28, colsample=1, and subsample=0.82 Moreover, for the Extra Trees model, we utilized the default hyperparameters. Since the primary objective of this study is not to identify the best-performing AI models, we focus on a basic implementation that can perform consistently across all datasets.

\section{Experimental Result and Discussions} 

In this section, we present and analyze the performance of our proposed models. We also introduce the performance of various classifiers across different datasets and propose an explainable framework. Our approach not only utilizes machine learning models to analyze patterns in encrypted traffic but also provides both global and local explanations for the best-performing algorithm.

\subsection{Detection Model Performance}
\begin{figure}[htbp]
    \centering
    \subfigure[]{%
        \includegraphics[width=0.3\textwidth]{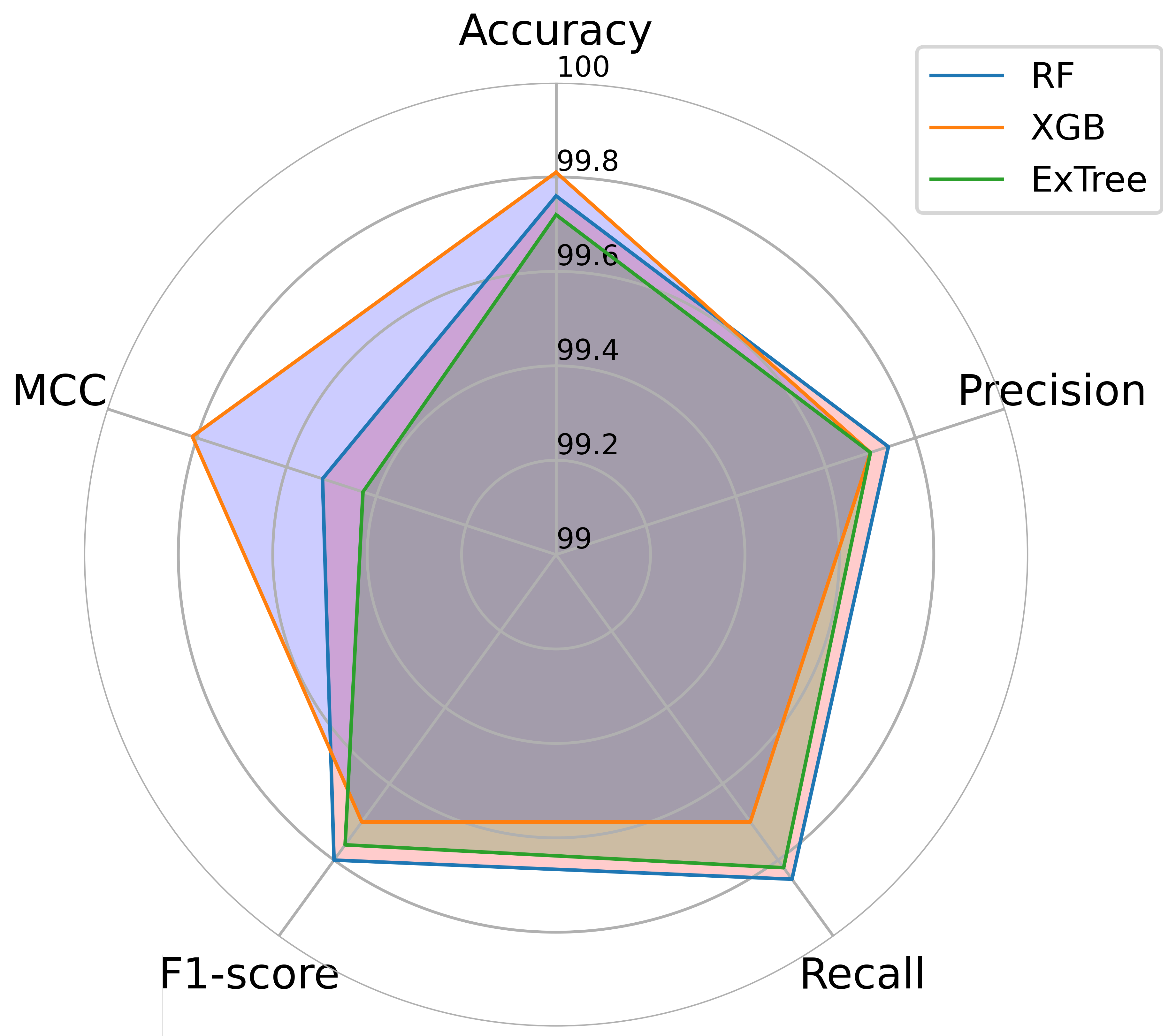}
    }
    \subfigure[]{%
        \includegraphics[width=0.3\textwidth]{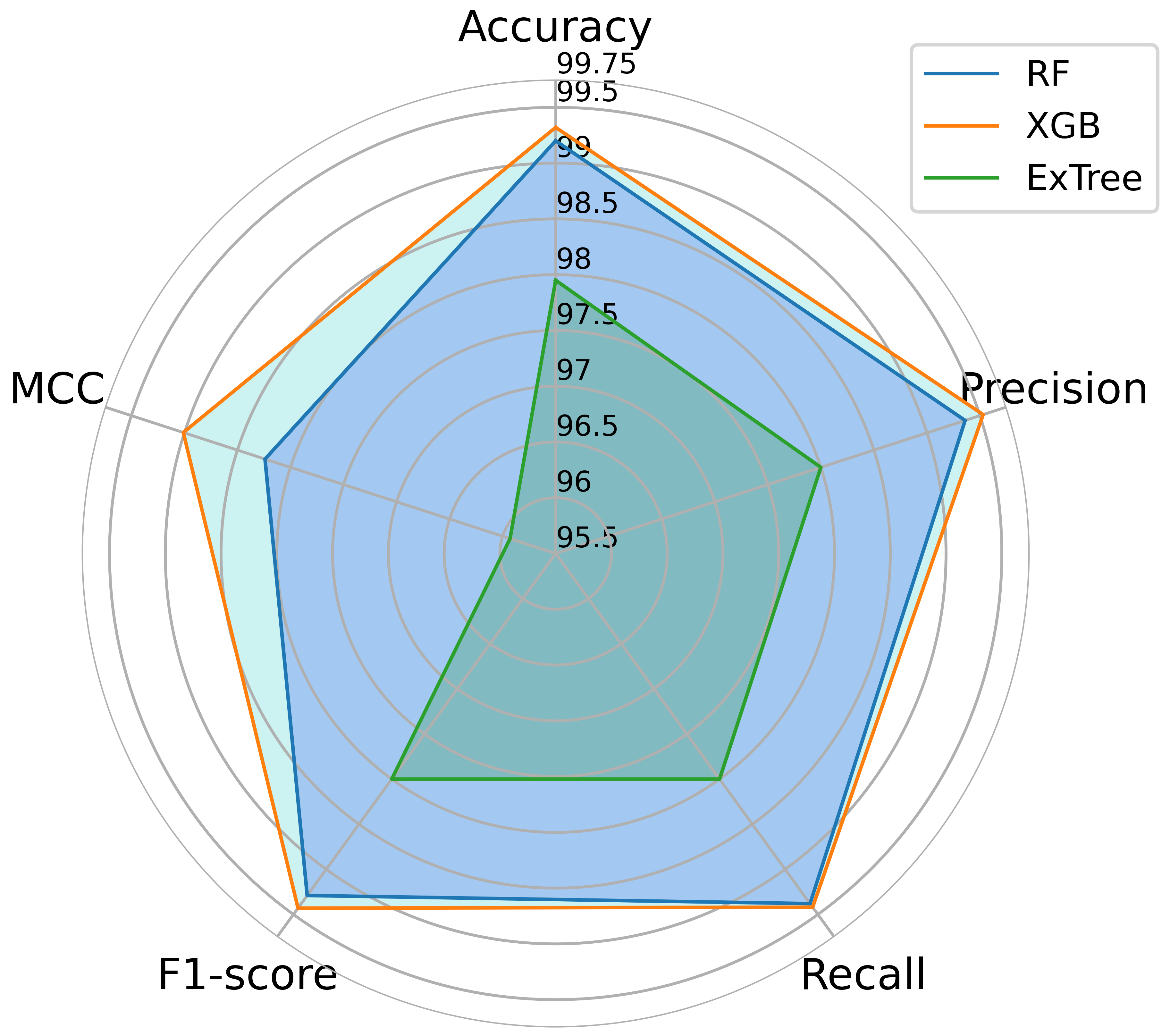}
    } 
    \subfigure[]{%
        \includegraphics[width=0.3\textwidth]{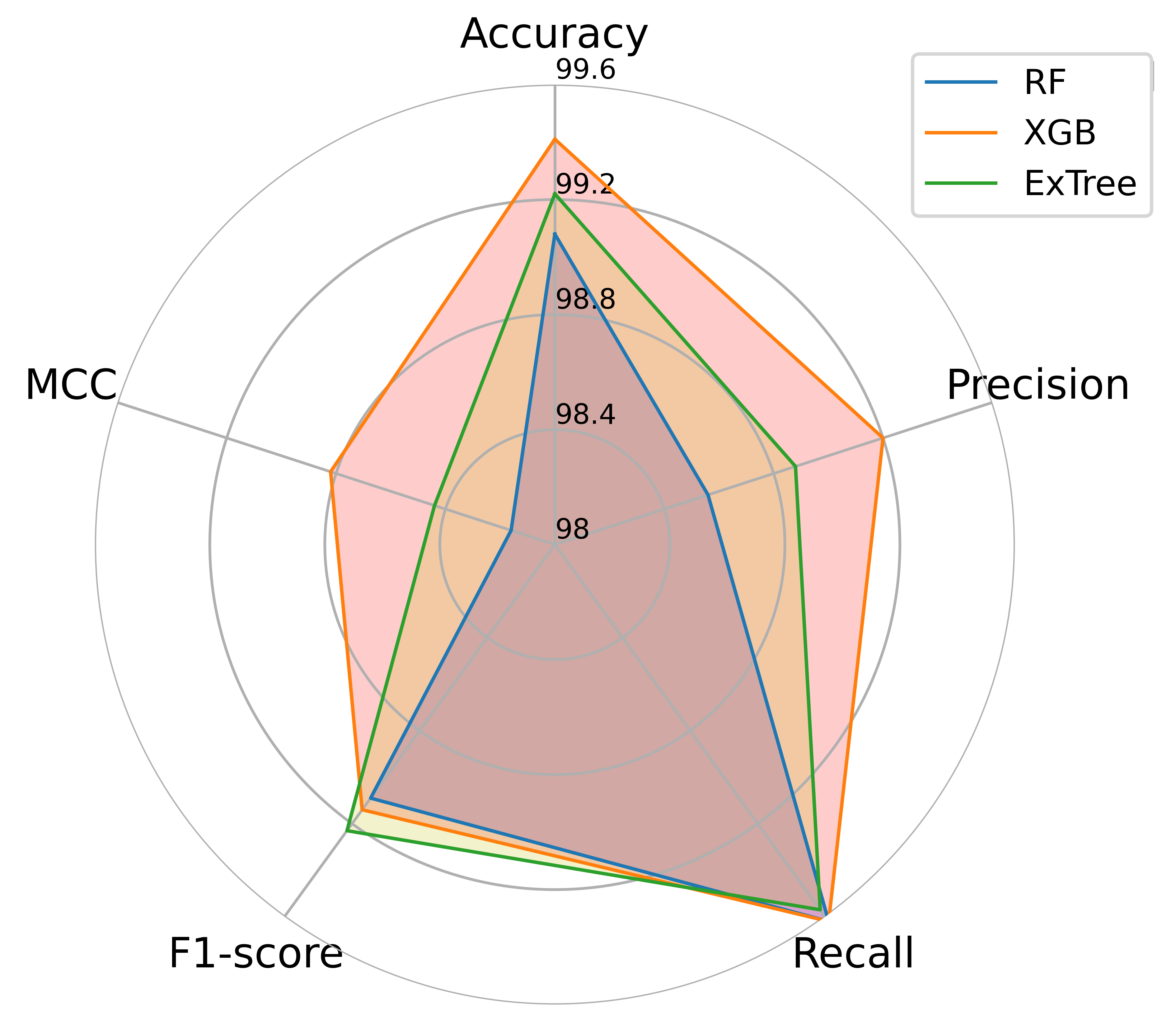}
    }
    \caption{Radar plot for performance comparison of the proposed malware detection  across different datasets and metrics: (a) CTU-13; (b) Our dataset; (c) Our imbalanced dataset}
    \label{performancePLt}
\end{figure}

The evaluation of the classification models on the CTU-13 dataset, our dataset (balanced), and imbalanced classes reveals key insights into their performance. As the CTU-13 dataset was collected in a real-world communication setting, the approaches and results can be replicated in other datasets and real-world scenarios. Across all datasets, XGB consistently achieved the highest MCC, making it the most reliable model for class prediction, as shown in Fig. \ref{performancePLt}. In the CTU-13 dataset, XGB attains an MCC of 99.81\%, surpassing RF (99.52\%) and ExTree (99.43\%), indicating superior overall predictive power, while effectively accounting for both false positives and false negatives, as shown in Fig.  \ref{performancePLt}(a). Similarly, in the balanced dataset shown in Fig.  \ref{performancePLt}(b), XGB again outperforms the other models with an MCC of 99.01\%, compared to RF’s 98.24\% and ExTree’s 95.93\%.

In the imbalanced version of our dataset, Fig. 2(c) shows that XGB maintains its edge with an MCC of 98.82\%, demonstrating robust performance in imbalanced settings compared to RF (98.16\%) and ExTree (98.44\%). The high MCC values for XGB across all datasets indicate that it is less influenced by skewed class distributions and manages true negatives more effectively than RF and ExTree, which are crucial for balanced evaluation. Additionally, the XGBoost model’s ability to handle missing values and its support for parallel processing makes it suitable for large-scale malware detection tasks.
 

\begin{figure}[htbp]
    \centering
    \subfigure[]{%
        \includegraphics[width=0.3\textwidth, height=80pt]{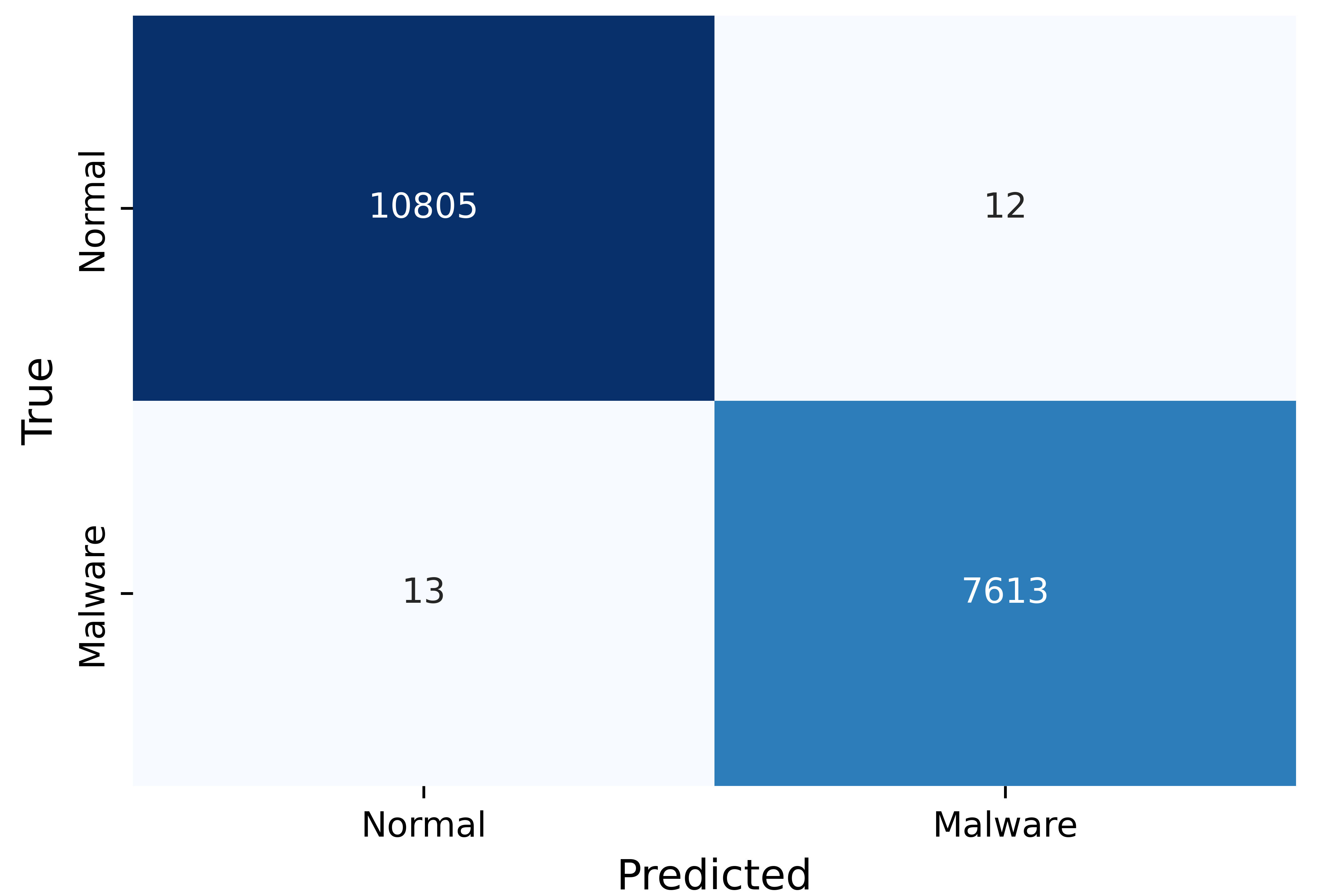}
    }
    \subfigure[]{%
        \includegraphics[width=0.3\textwidth, height=80pt]{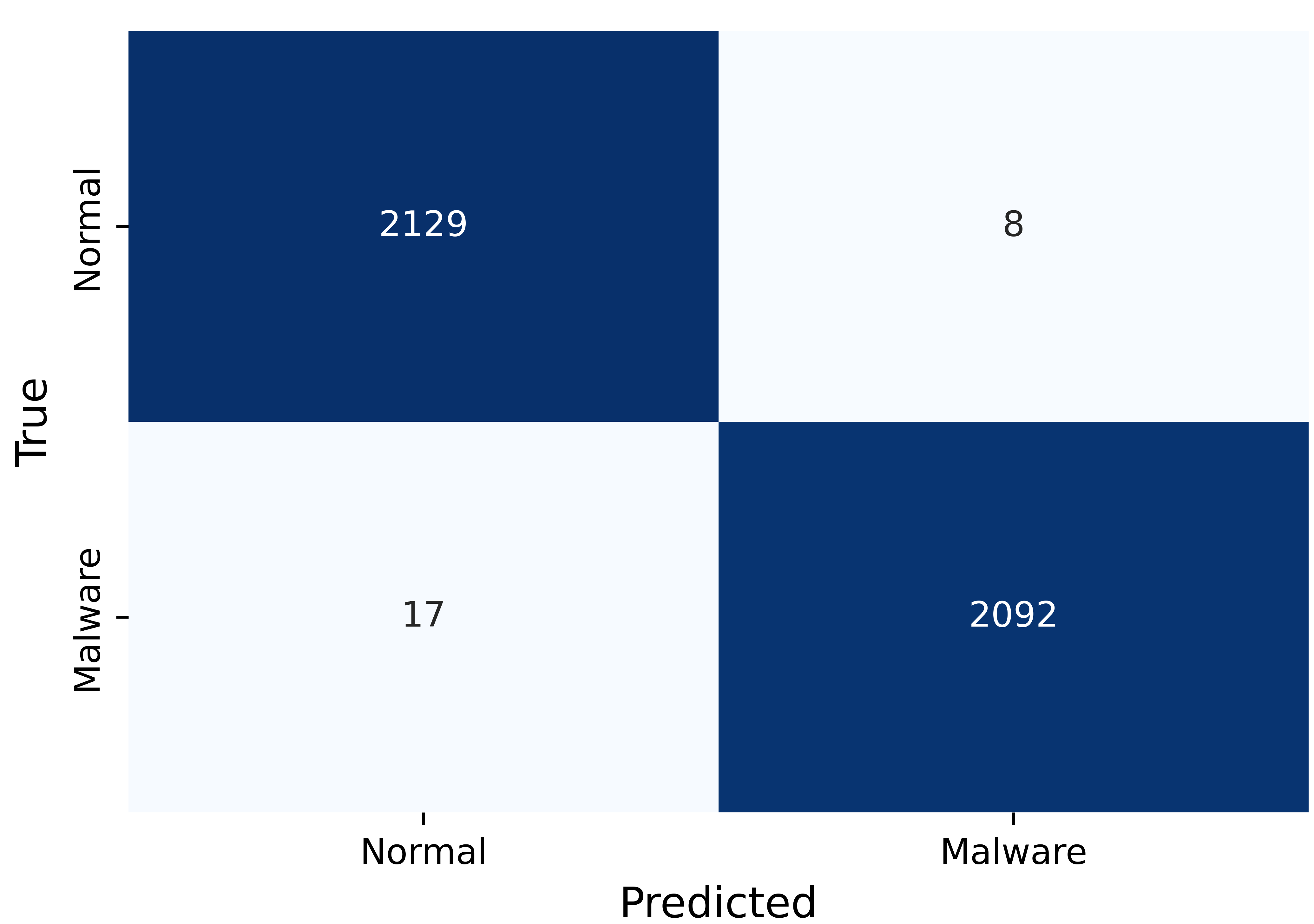}
    }
    \caption{Confusion matrix comparison on: (a) CTU-13; (b) Our imbalanced dataset}
    \label{confusion}
\end{figure}

The confusion matrix in Fig.  \ref{confusion} illustrates the ability of the model to accurately classify malicious and normal traffic. In the CTU-13 dataset, shown in Fig. \ref{confusion}(a), the model achieves near-perfect detection with minimal false positives and false negatives. Fig. \ref{confusion}(b) shows the performance based on our dataset trained using oversampling techniques. Although the false negative rate is slightly higher, the low false positive rate ensures that normal traffic is not wrongly flagged as malware, making it suitable for deployment in real-world encrypted network environments, where accuracy is paramount. The slightly higher false negative count may be attributed to variations in the traffic patterns or encryption methods. However, the model continues to demonstrate robust generalization to new data. 

In both cases, the model’s robust performance indicates its capability to manage the complexities introduced by encryption, which has traditionally posed challenges for malware detection owing to limited visibility in the content of network packets. The low rates of both false positives and false negatives imply that this approach can enhance the precision and effectiveness of malware detection systems, in an encrypted environment.

\subsection{Explaining Detection Model}
Visualizing the best-performing model’s decision-making process using XAI techniques helps to identify malicious behavior and understanding such a decision is crucial for cyber-analytics. An insight into the decision helps administrators identify the part of the network, part of the features, the security policies compromised by attackers, and potential biases using the techniques provided by SHAP.

\subsubsection{Global Explainability:} 
\begin{figure}[H] 
\centering
\vspace{-8mm}

\includegraphics[width=0.95 \textwidth ]{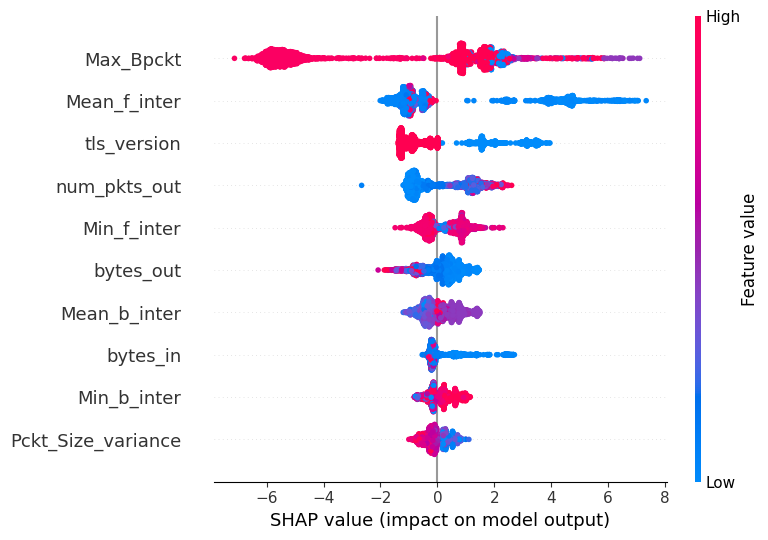}
\caption{Summary plot of global explanation of XGBoost model trained using our dataset} \label{fig2}
\end{figure}

The SHAP summary plot illustrates the global importance of features, showing their contribution to model decisions and their impact direction (positive or negative) for the individual samples. Fig. \ref{fig2} presents the ten most influential features of the XGBoost model trained on our dataset. The color bar indicates the impact of the features, with blue indicating positive influences and red indicating negative ones. The horizontal violin plot displays the distribution of Shapley values for each data instance. A higher ‘Max\_Bpckt’ value significantly impacts decision-making, distinguishing between malicious and normal traffic based on packet size. The ‘Mean\_f\_inter’ feature captures the average time interval between consecutive packets during a session, as malware traffic often exhibits distinctive timing characteristics compared to legitimate traffic, such as irregular or bursty patterns. Additionally, feature bytes\_in, which is the number of bytes sent from the client, positively impacts the model's decision, revealing the distinct behavior of malicious servers that exfiltrate client data. This finding regarding feature importance corroborates a previous study on adversarial malicious encrypted malware traffic analysis \cite{30}.

\begin{figure}
\centering 
\includegraphics[width=  0.95\textwidth ]{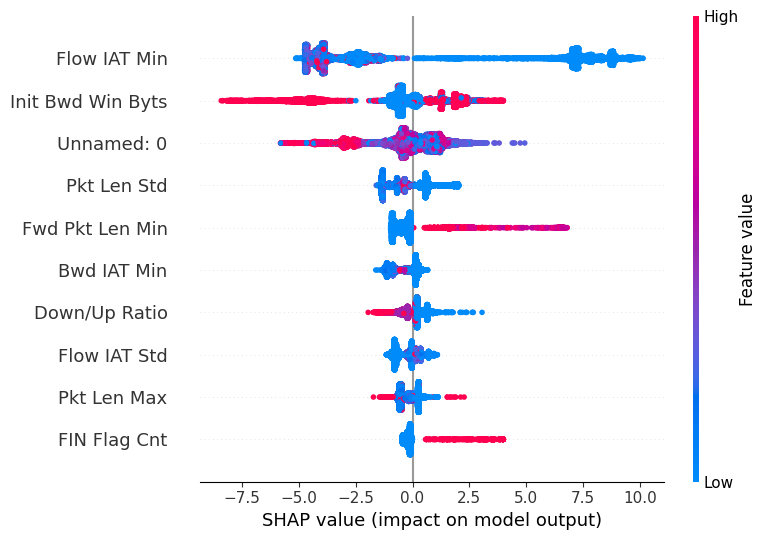}
\caption{Summary plot of global explanation of XGBoost model trained using CTU-13 dataset} \label{fig3}
\end{figure}
In normal traffic, the standard deviation of the byte inter-arrival time between the two flows positively impacts decision-making. Additionally, 'certValidDays', which represents the number of days a certificate remains valid, significantly influenced the results. Malware authors aim to remain undetected, leading to a higher number of validation days in the malware communication. Fig. \ref{fig3} illustrates the model trained on CTU-13 data, revealing that features like minimum forward packet length 'fwd Pkt Len Min', packet length standard deviation 'Pkt Len STD' and bytes forwarded 'init Bwd Win Bytes' affect the models positively impact model performance like our dataset.
 
\subsubsection{Local Explainability:}

\begin{figure}
\centering 
\includegraphics[width=\textwidth, height=90pt]{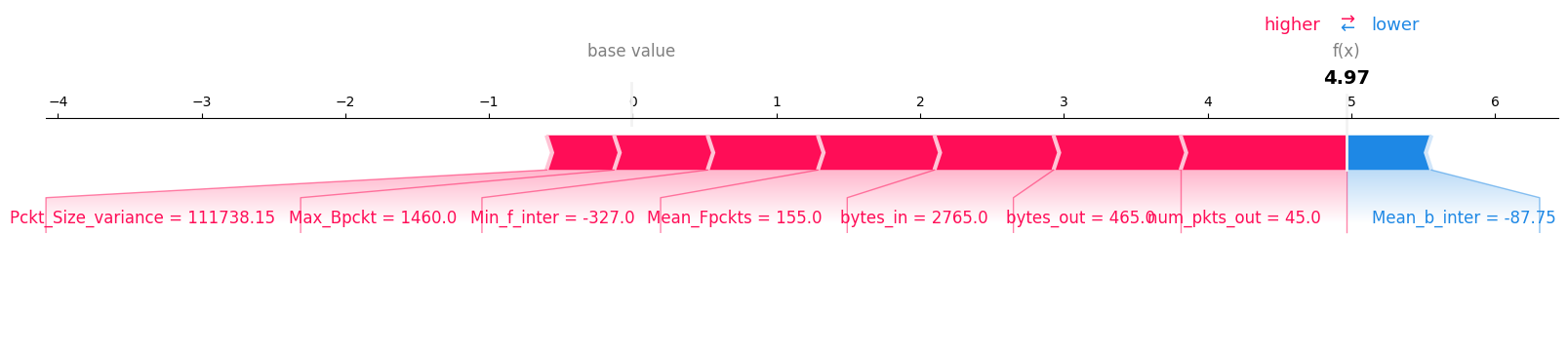}  
\caption{Local explanation based on our dataset} 
\label{fig4} 
\end{figure}

\begin{figure}
\centering 
\includegraphics[width=\textwidth, height=95pt]{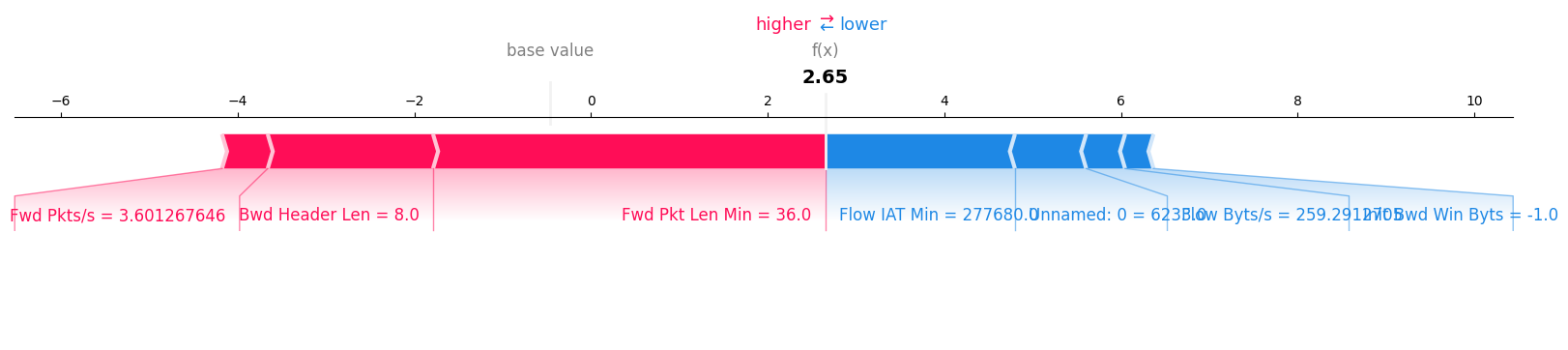}  
\caption{Local explanation based on CTU-13 dataset}
\label{CTUFig}
\end{figure}

The force plot is a part of TreeExplainer in SHAP. It is useful for visualizing the impact of features on individual predictions made by tree-based models. Fig. \ref{fig4}  and \ref{CTUFig} show the model’s prediction decomposed into the sum of the effects of each feature value on the model output from the base value prediction. The explanation of an expected feature that affects the target class centers on the plot around the x-axis. Features that have a favorable influence on the prediction are displayed in red, whereas those that have a negative influence are displayed in blue. As shown in Fig. \ref{fig4}, was 4.97, which is higher than the base value indicating that the features associated with the malware class are pushed to the right (higher values) as predicted by the red color. For instance, 'num\_pkts\_out' positively affects the malware class (indicated by the red region), thereby shifting the class malware prediction to the right. In contrast, the malware class is negatively correlated with the feature 'Flow IAT Min’ feature, as shown in Fig. \ref{CTUFig}. A lower value indicates that malware prediction is pushed to the left. However, the force to drive the prediction to the right (higher value) is the largest because the feature values that are in the red area are greater than the other feature values that are in the blue area.   

\subsection{Discussion}

The global explanation reveals that the mean number of bytes sent and received by a client and server indicates a significant difference in the data transmission patterns between normal traffic and that associated with malware. Specifically, the data transmitted from the server to the client are typically greater in normal traffic than in malware scenarios. This suggests that legitimate client applications tend to receive a substantial amount of data at the onset of communication, in contrast to malware client applications. Conversely, when a client transmits data to the server, the mean number of bytes in the normal traffic is comparatively lower. This observation may reflect the behavior of malware, which often involves collecting and transmitting a client's status and data to a command and control server. Furthermore, the model's focus on the packets sent and received by clients and servers indicates that malware traffic is characterized by a higher volume of packets transferred. This pattern suggests that malware actors deliberately send numerous packets containing small amounts of data to evade detection, thereby minimizing bandwidth consumption and reducing the likelihood of triggering security alerts.

 Furthermore, the TLS employed in the communication provided critical insights. As demonstrated in Fig. \ref{fig2} and \ref{CTUFig}, the TLS version significantly influences the performance of the detection model. This phenomenon can be attributed to the tendency of normal communications to favor newer TLS versions, whereas malicious actors often opt for older versions. Generally, when selecting cipher suites, malware authors tend to prefer simpler and outdated parameters to optimize the utilization of limited computational resources. In addition, the inter-arrival time differences between multiple packets within a flow are utilized as time-series elements. The interval times of forwarded backward flows can substantially impact models, with normal flows exhibiting a more positive correlation than those associated with malware. By identifying the most significant features for detection, security professionals can prioritize their efforts more effectively. Moreover, insights derived from SHAP values can assist in the customization of alert systems. For instance, if communication to a web server indicates the transmission of a large packet, it may raise concerns regarding the socket, potentially triggering an alert..
\subsection{Comparison of the Proposed Work with Existing Works}
In this section, we present a comparative analysis of our approach to explainable malware detection in encrypted network traffic, about existing methodologies. By evaluating both traditional and contemporary methods, we intend to underscore the innovations and enhancements presented by our model, particularly regarding its capacity to provide transparency in decision-making while sustaining robust performance in complex encrypted environments. The proposed explainable model incorporates TLS, statistical, and flow metadata features for binary classification, resulting in an innovative approach to encrypted malware detection. Moreover, \cite{38} proposed an ensemble model using self-attention techniques for multiclass classification and achieved an accuracy of 96.71\%. Our system outperforms in terms of accuracy metrics when using the CTU-13 dataset in addition to the enhancement of the interpretability of models.

\section{Conclusion} 
Encrypted network communications play a critical role in safeguarding privacy. The ability to detect malware without decrypting the payload while providing explanations for detection decisions is a vital practice in the field of cybersecurity. This process requires the development and evaluation of features that accurately represent encrypted traffic through the application of artificial intelligence algorithms. The extracted features encompass a range of attributes associated with the encrypted communication between clients and servers. These attributes include handshake details, such as the ciphers offered and accepted, the TLS extensions that are advertised and supported, certificate information, timing and length characteristics, connection metadata, and statistical data regarding packets and bytes in bi-directional flow. Although several studies have focused on malware detection, there is a notable scarcity of research addressing the explainability of such detection systems. In this study, we introduce an efficient explainable detection model based on tree ensemble methods. The proposed framework was evaluated using both a self-collected dataset and a baseline dataset. Our detection results demonstrated significant improvements compared to those of related studies. 

 The primary objective of this study is to investigate the explainability of a model's decisions. We utilized the SHAP method to examine both global and local explainability, thereby identifying the features that exerted the greatest influence on the model's predictions. The results illustrate the varying contributions of the different features to the model for each class.  Moreover, we also contribute a comprehensive dataset composed of 54 different malware types and 1,127 unique malware traffic captures that occurred in cyberspace. In future research, a more comprehensive analysis of malware communication, along with enhanced explainability through the use of additional samples to detect zero-day attacks, may represent a promising avenue for exploration.

\subsubsection{\ackname} This study is partially supported by the Age-It project under the National Recovery and Resilience Plan (NRRP) program funded by the NextGenerationEU.

\bibliographystyle{splncs04}

\end{document}